\documentclass[twocolumn,twoside,slac_two]{revtex4}
\usepackage{graphicx}
\usepackage{fancyhdr}
\newcommand{\epoe}{\epsilon '/\epsilon}
\newcommand{\reepoe}{Re(\epoe)}

\newcommand{\chg}{\pi^+ \pi^-}
\newcommand{\neut}{\pi^0 \pi^0}

\begin{document}

\title{Review on $\epsilon '/\epsilon$}

%

\author{Taku Yamanaka}
\affiliation{Osaka University, Osaka, Japan}

\begin{abstract}
Experiments at CERN and Fermilab have been competing each other to 
improve the measurement of a CP violation parameter, 
$\reepoe$.
Fermilab KTeV-E832 recently announced their final result, 
\(\reepoe = [19.2 \pm 2.1] \times 10^{-4}\).
The new world average shows the existence of direct CP violation
in the decay process itself with 12 standard deviations.
\end{abstract}

\maketitle

\thispagestyle{fancy}


\section{Introduction}
Violation of CP conservation was first observed in the 
\(K_L \to \pi^+ \pi^-\) decay \cite{cronin},
in which CP ``odd" $K_L$ was decaying into CP even $\pi^+ \pi^-$ state.
This phenomena is referred to as 
{\em indirect} CP violation, 
because it is caused by 
an imaginary phase in the 
\(K^0 - \bar{K}^0\) (\(\Delta S = \pm 2\))
transition amplitude.
This phase adds 
small amount of CP even $K_1$ state to the CP odd $K_2$ state 
to construct the $K_L$ state:
\(|K_L> \simeq |K_2> + \epsilon |K_1>\).
Among the many theoretical models that tried to explain the source of 
the imaginary phase, two survived for many years.
The superweak model \cite{sw} postulated that there was
a new \(\Delta S = 2\) interaction that
introduced the imaginary phase in the 
\(K^0 - \bar{K}^0\) mixing.
Kobayashi and Maskawa pointed out that the imaginary phase is introduced 
naturally in a mixing between three generations of quarks\cite{km}.
This scheme became part of the standard model, wherein 
{\em indirect} CP violation is explained by a top quark contribution in a 
\(K^0 - \bar{K}^0\) box diagram.
The standard model also predicts that CP violation can occur in 
$\Delta S = 1$ decay processes through a penguin diagram.
This makes a direct transition from CP odd $K_2$ state to a CP even 
final state.
This is called {\em direct} CP violation, and its
size is expressed by the model-independent parameter $\epsilon '$.
The Superweak model cannot produce {\em direct} CP violation
because the decay itself is not a \(\Delta S = 2\) process.
Therefore, measurement of non-zero $\reepoe$ became an important 
experimental focus.  If it is not zero,
{\em direct} CP violation exists and the Superweak model is falsified.

The parameter, $\reepoe$, is determined by measuring the ratio of 
partial decay widths, 
\begin{eqnarray}
R & = & \frac{\Gamma(K_L \to \chg)/\Gamma(K_S \to \chg)}
	{\Gamma(K_L \to \neut)/\Gamma(K_S \to \neut)}\\
& = & 1 + 6 Re(\epsilon '/\epsilon).
\end{eqnarray}
Since the required accuracy on $R$ is \(O(10^{-3})\), 
systematic errors had to be controlled to a smaller level,
which was challenging experimentally.

\section{Past Results}
There have been two major efforts in the world; 
one at CERN and one at Fermilab,
competing each other to make precise and accurate measurements.

\subsection{CERN NA31 and NA48 Experiments}
The original experiment at CERN, NA31\cite{na31}, had one kaon beam, 
and had $K_L$ runs with a target far upstream, and
separate $K_S$ runs with a production target moving in the decay region.
In this scheme, the detector rates were not exactly the same between
$K_L$ and $K_S$ runs.

\begin{figure}[htbp]
	\centering
	\includegraphics[width=80mm,angle=90]{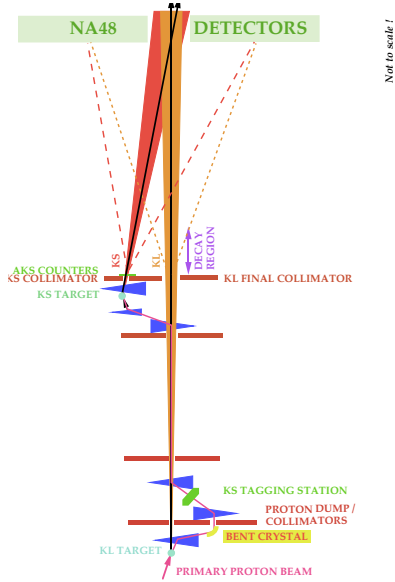}
	\caption{CERN NA48 beamline.  
		$K_L$ and $K_S$ were produced at different targets, 
		and the two beams merged at the detector region.} 
	\label{fig:na48_beam}
\end{figure}

The improved experiment at CERN, NA48, used two production targets to 
make $K_L$ and $K_S$ beams simultaneously, as shown in Fig.~\ref{fig:na48_beam}.
The two beams merged at the detector region, and $K_S$ decays were
identified by protons passing through a set of tagging counters 
just before hitting the $K_S$ production target.
For $\chg$ decays, the decay vertex and the momenta of the 
pions were measured with 
four sets of drift chambers and a spectrometer magnet.
For $\neut$ decays, the energies and hit positions 
of photons were measured by a liquid krypton calorimeter.
The decay vertex position was reconstructed by assuming 
the kaon mass for the four photon invariant mass.
The reconstructed decay vertex position distribution for 
$K_L$ decays were weighted to have the same distribution as
$K_S$ decays, to reduce systematic errors.
Their final result \cite{na48} based on their all data sets is
\(\reepoe = [14.7 \pm 2.2] \times 10^{-4}\),
showing a clear deviation from zero.

\subsection{Fermilab E731 and KTeV-E832 Experiments}
The E731 experiment\cite{e731} at Fermilab used a $K_L$ beam and 
a regenerated $K_S$ beam, and observed the four modes simultaneously.
Its sensitivity was limited by the beam intensity and performance of
the lead glass electromagnetic calorimeter.

To run at higher beam intensity and to improve systematic errors, 
the KTeV-E832 experiment was built with a completely new beam line and detector.
The collaboration had two runs, and their result \cite{ktev97} from the first 
run in 1997,
\(\reepoe = [20.7 \pm 1.48 (stat.) \pm 2.39 (syst.)] \times 10^{-4}\),
showed a 7$\sigma$ deviation from zero.
Since then, they have improved their simulation and analysis, and have 
recently announced their new and final result based on the full data set.
 
\section{New and Final Results from Fermilab KTeV}
\begin{figure}[htbp]
	\centering
	\includegraphics[width=80mm]{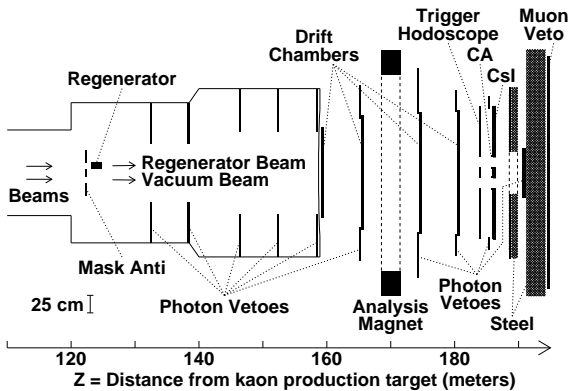}
	\caption{Plan view of the KTeV-E832 experiment.} 
	\label{fig:ktev_det}
\end{figure}

Figure \ref{fig:ktev_det} shows the plan view of the KTeV-E832 experiment.
The experiment used two nearly parallel $K_L$ beams, 
and placed a 1.68m long scintillator blocks in one of the beams 
to regenerate $K_S$.  
Requiring that there be no energy deposit in the scintillator block 
selected coherently regenerated $K_S$ with has the same 
beam divergence as $K_L$.
The momenta of $\chg$ tracks were measured with four sets of drift chambers
and a dipole magnet.
The four photons from $\neut$ decays were measured with 
an electromagnetic calorimeter made of pure CsI crystals.

\begin{figure}[htbp]
	\centering
	\includegraphics[width=80mm]{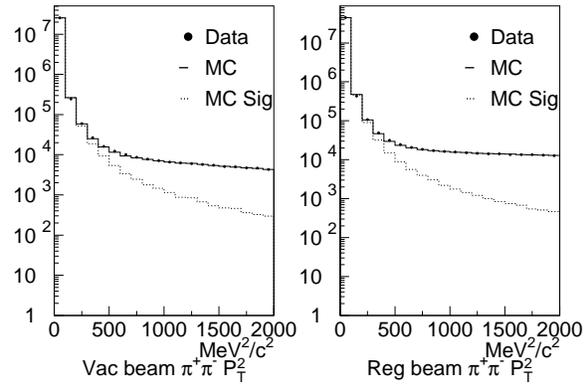}
	\caption{The distribution of the square of the transverse momentum of 
		the $\chg$ system, $P_T^2$,
		for left:vacuum ($K_L$) and right:regenerator ($K_S$) beams.  
		The dots show data, and the histogram (MC) shows Monte Carlo simulation
		for signal plus background.} 
	\label{fig:ktev_pt2}
\end{figure}

There were many improvements on the Monte Carlo simulation and data analysis.
For example, for $\chg$ decay mode, 
the chamber resolution was measured as a function of 
the position within the drift cell 
and this measurement was used in Monte Carlo simulation and
track resonstruction.
In addition, the simulation modeled
$\delta$-rays crossing multiple cells in the chamber, 
bremstrahlung downstream of the magnet, 
hadronic interactions, and 
$dE/dx$ in materials ($\sim$4.5MeV).
As shown in Fig.~\ref{fig:ktev_pt2},
all these small improvements made 
the distribution of transverse momentum of the 2 track system agree better 
between data and Monte Carlo simulation, 
reducing the corresponding systematic uncertainty from 
\(0.25 \times 10^{-4}\) in the previous result to 
\(0.10 \times 10^{-4}\).

\begin{figure}[htbp]
	\centering
	\includegraphics[width=80mm]{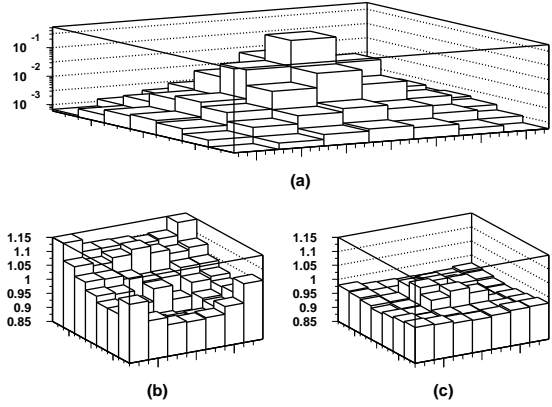}
	\caption{(a): The fraction of energy in each of the $7 \times 7$ CsI crystals
		in an electron shower for data.
		(b): The data/MC ratio for the 2003 paper, and 
		(c): the new analysis.} 
	\label{fig:shower}
\end{figure}

For the $\neut$ mode, the light uniformity and non-linearity of the CsI were
corrected for in each crystal.
Electromagnetic showers were simulated with finite incident angles, 
and 13$\mu m$ thick Aluminized Mylar wrappings and shims between crystals
were also added to the simulation.
With these improvements, Monte Carlo simulation reproduced the shower shape 
better, as shown in Fig.~\ref{fig:shower}.
The reconstructed kaon mass dependence on the photon incident angle and kaon 
energy also agree better between data and Monte Carlo simulation.
These better agreements reduced the corresponding systematic uncertainty 
from \(1.47 \times 10^{-4}\) to
\(0.75 \times 10^{-4}\).

\begin{figure}[htbp]
	\centering
	\includegraphics[width=80mm]{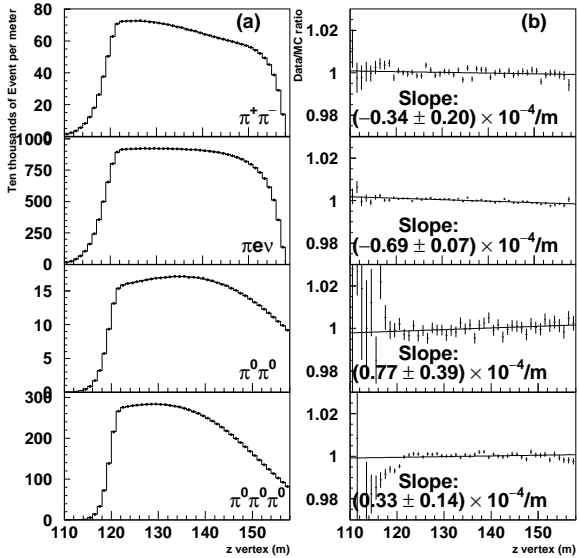}
	\caption{(a): From top to bottom, the decay vertex distributions for 
		\(K_L \to \chg\), \(K_L \to \pi e \nu\), 
		\(K_L \to \neut\), and \(K_L \to \pi^0 \pi^0 \pi^0\)
		decays.
		(b): The data to MC ratios as a function of the decay vertex position
		for each decay mode.
		} 
	\label{fig:zdist}
\end{figure}

In addition to the data from the 1999 run, KTeV-E832
also reanalyzed the data from the 1997 run.
The numbers of observed events after event selections were:
25.1M for \(K_L \to \chg\), 43.7M for \(K_S \to \chg\), 
6.0M for \(K_L \to \neut\), and 10.2M for \(K_S \to \neut\).
The numbers of actual decays in 10 GeV/c kaon momentum bins were
calculated by correcting for the acceptance.
The acceptance was determined by Monte Carlo simulation, and it was
checked with high statistics decay modes, such as \(K_L \to \pi^\pm e^\mp \nu\) and 
\(K_L \to \pi^0 \pi^0 \pi^0\), as shown in Fig.~\ref{fig:zdist}.

The major systematic uncertainties are
\(1.07 \times 10^{-4}\) for background estimation in $\neut$ modes,
\(0.75 \times 10^{-4}\) for CsI cluster reconstruction, 
\(0.57 \times 10^{-4}\) for $\chg$ mode acceptance,
\(0.48 \times 10^{-4}\) for $\neut$ mode acceptance, 
\(0.48 \times 10^{-4}\) for detector apertures in $\neut$ mode, etc..
The total systematic error was reduced from \(2.39 \times 10^{-4}\) to 
\(1.78 \times 10^{-4}\).

The final result on the full KTeV data is 
\(\reepoe = [19.2 \pm 1.1 (stat.) \pm 1.8 (syst.)] \times 10^{-4}
= [19.2 \pm 2.1] \times 10^{-4}\).

Using the same data, KTeV-E832 significantly improved other kaon parameter measurements,
\(\Delta m = (5265 \pm 10) \times 10^6 \hbar/s\),
\(\tau_S = (89.62 \pm 0.06) \times 10^{-12} s\),
\(\phi_\pm = \arg(\eta_\pm) = (44.1 \pm 1.0)^\circ\), and
\(\Delta \phi = -3Im(\epoe) = (0.30 \pm 0.35)^\circ\).


\section{Conclusion}
\begin{figure}[htbp]
	\centering
	\includegraphics[width=80mm]{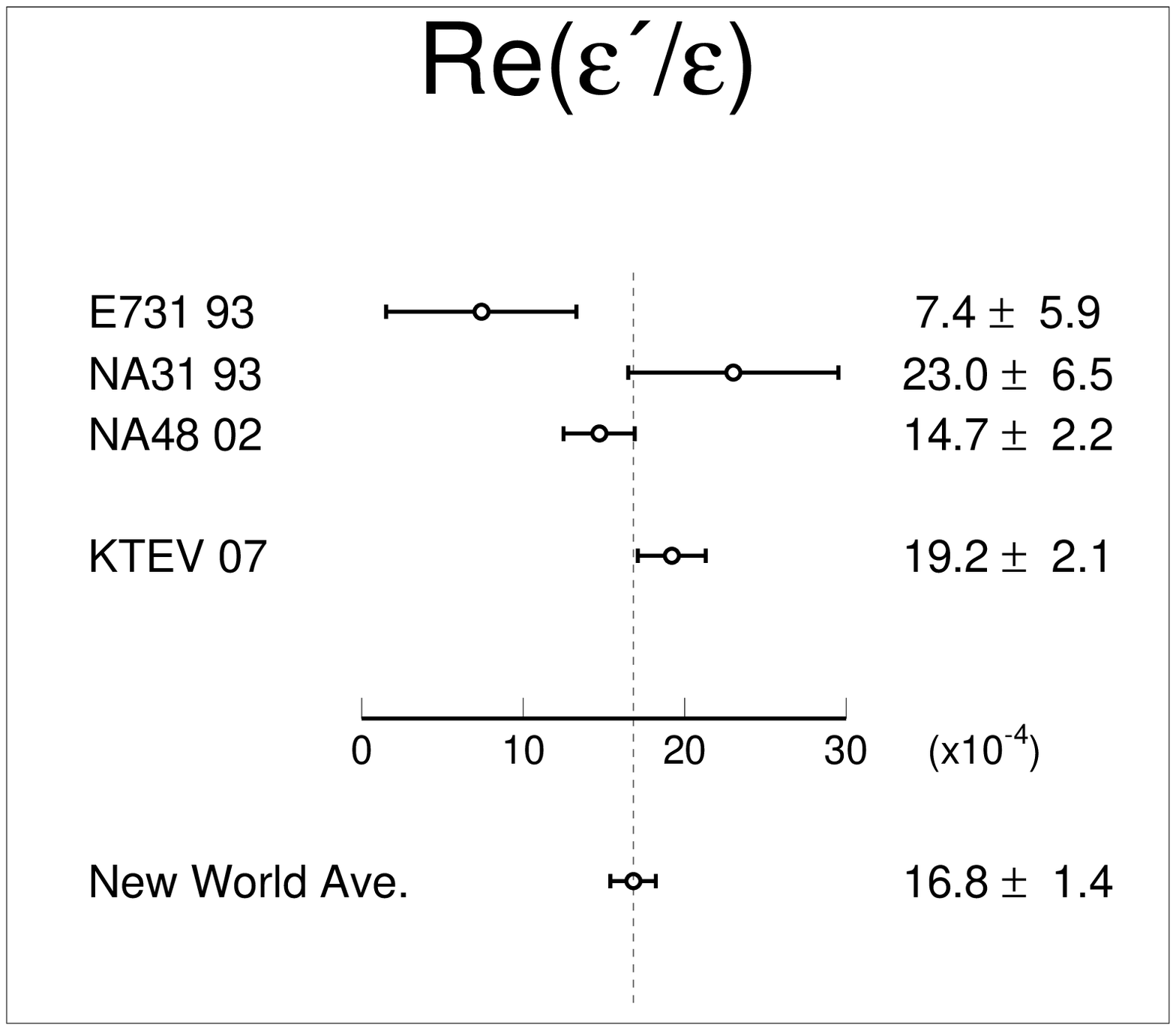}
	\caption{History of $\reepoe$ measurements.} \label{fig:history}
\end{figure}

Figure \ref{fig:history} summarizes the results of the $\reepoe$ measurements.
Combining the new KTeV result with the past results, 
the new world average on $\reepoe$ is 
\([16.8 \pm 1.4] \times 10^{-4}\).
The CERN and Fermilab experiments have clearly established 
the existence of direct CP violation and rejected the Superweak model.

\bigskip 

\begin{thebibliography}{9}   
	\bibitem{cronin} J.H.~Christensen {\it et al.}, Phys. Rev. Lett. {\bf 13}, 138 (1964).
	\bibitem{sw} L. Wolfenstein, Phys. Rev. Lett. {\bf 13}, 562 (1964).
	\bibitem{km} M.~Kobayashi and K.~Maskawa, Prog. Theor. Phys. {\bf 49}, 652 (1973).
	\bibitem{na31} G.D.~Barr {\it et al.}, Phys. Lett. B {\bf 317}, 233 (1993).
	\bibitem{e731} L.K.~Gibbons {\it et al.},Phys. Rev. Lett. {\bf 70}, 1203 (1993). 
	\bibitem{na48} J.R.~Batley {\it et al.}, Phys. Lett. B {\bf 544}, 97 (2002).
	\bibitem{ktev97} A.~Alavi-Harati {\it et al.}, Phys. Rev. D {\bf 67}, 012005 (2003).
\end{thebibliography}

\end{document}